\documentclass[iop,apj,tighten]{emulateapj}

\submitted{Received 2016 June 29; revised 2016 September 14; accepted 2016 September 15}
\journalinfo{The Astrophysical Journal}

\usepackage{apjfonts} 
\usepackage{amsmath,amstext}
\usepackage[breaklinks,colorlinks,citecolor=black,linkcolor=black,urlcolor=black]{hyperref} 
\usepackage[all]{hypcap} 

\newcommand{\Ni}{\ensuremath{^{56}\mathrm{Ni}}}
\newcommand{\Co}{\ensuremath{^{56}\mathrm{Co}}}

\newcommand{\Mej}{\ensuremath{M_\mathrm{ej}}}
\newcommand{\Mni}{\ensuremath{M_\mathrm{\Ni}}}
\newcommand{\Eej}{\ensuremath{E_\mathrm{ej}}}
\newcommand{\Ecoll}{\ensuremath{E_\mathrm{coll}}}
\newcommand{\Msun}{\ensuremath{M_\odot}}

\newcommand{\kmps}{\ensuremath{\mathrm{km~s^{-1}}}}
\newcommand{\tbh}{\ensuremath{t_\mathrm{BH}}}
\newcommand{\ej}{\ensuremath{\mathrm{ej}}}

\shorttitle{SNe powered by magnetars that transform into BHs}
\shortauthors{Moriya, Metzger, \& Blinnikov}

\begin{document}

\title{Supernovae powered by magnetars that transform into black holes}
\author{Takashi J. Moriya}
\affil{Division of Theoretical Astronomy, National Astronomical Observatory of Japan, National Institues of Natural Sciences, \\ 2-21-1 Osawa, Mitaka, Tokyo 181-8588, Japan \\ takashi.moriya@nao.ac.jp}

\author{Brian D. Metzger}
\affil{Columbia Astrophysics Laboratory, Columbia University, New York, NY 10027, USA}

\author{Sergei I. Blinnikov}
\affil{Institute for Theoretical and Experimental Physics, Bolshaya Cheremushkinskaya ulitsa 25, 117218 Moscow, Russia}
\affil{All-Russia Research Institute of Automatics, Sushchevskaya ulitsa 22, 127055 Moscow, Russia}
\affil{Kavli Institute for the Physics and Mathematics of the Universe (WPI), The University of Tokyo Institutes for Advanced Study, The University of Tokyo, \\ 5-1-5 Kashiwanoha, Kashiwa 277-8583, Japan}

\begin{abstract}
Rapidly rotating, strongly magnetized neutron stars (magnetars) can release their enormous rotational energy via magnetic spin-down, providing a power source for bright transients such as superluminous supernovae. On the other hand, particularly massive (so-called supramassive) neutron stars require a minimum rotation rate to support their mass against gravitational collapse, below which the neutron star collapses to a black hole.  We model the light curves of supernovae powered by magnetars which transform into black holes.  Although the peak luminosities can reach high values in the range of superluminous supernovae, their post maximum light curves can decline very rapidly because of the sudden loss of the central energy input.  Early black hole transformation also enhances the shock breakout signal from the magnetar-driven bubble relative to the main supernova peak.  Our synthetic light curves of supernovae powered by magnetars transforming to black holes are consistent with those of some rapidly evolving bright transients recently reported by \citet{arcavi2016}.
\end{abstract}

\maketitle

\section{Introduction}
Some neutron stars (NSs) formed during the core collapse of massive stars are suggested to rotate very rapidly and, possibly for the same reason, to be strongly magnetized (\citealt{duncan1992}; \citealt{moesta2015}).  These strongly magnetized, rapidly rotating NSs are often referred as ``millisecond magnetars,''  although their connection$-$if any$-$to the high energy Galactic transients known also as magnetars is presently unclear.  Millisecond magnetars possess a prodigious reservoir of rotational energy $\sim 10^{51}-10^{52}$ erg, which can be extracted during the first seconds to weeks after the explosion through electromagnetic dipole spin-down.  If the energy from the magnetar wind is efficiently thermalized behind the expanding supernova (SN) ejecta shell (\citealt{Metzger+14}, see also \citealt{badjin2016}), then the resulting power source can greatly enhance the luminosity of the SN \citep[e.g.,][]{ostriker1971,shklovskii1976,mazzali2006,maeda2007}.

Recent work on the magnetar model was motivated by the discovery of superluminous SNe (SLSNe)\footnote{In this paper, we focus on Type~Ic SLSNe and refer them as ``SLSNe''. Most Type~II SLSNe are Type~IIn SNe and their power source is likely the interaction between SN ejecta and dense circumstellar media (e.g., \citealt{moriya2013}, but see also \citealt{inserra2016}).}, events with peak luminosities greater than $\sim 10^{44}~\mathrm{erg~s^{-1}}$, i.e., more than an order of magnitude brighter than typical core-collapse SNe \citep{quimby2011,gal-yam2012}.  For a surface magnetic dipole field strength of $\sim 10^{14}~\mathrm{G}$ and an initial rotational period of $\sim$ few $\mathrm{ms}$, the magnetar releases sufficient rotational energy ($\gtrsim 10^{51}~\mathrm{erg}$) over a proper timescale ($\sim 1-10$~days) to power SLSNe \citep{kasen2010}.  Although magnetars provide one possible explanation for SLSNe \citep[e.g.,][]{kasen2010,woosley2010,dessart2012,inserra2013,Metzger+14,bersten2016,sukhbold2016,mazzali2016}, several alternative models are also actively being explored \citep[e.g.,][]{moriya2010,chevalier2011,kasen2011,chatzopoulos2012b,dessart2013,kozyreva2014,sorokina2015}.  

The growing number of well-sampled SLSN light curves (LCs) has revealed a rich diversity of behaviors.  Some, and possibly all \citep{nicholl2016}, SLSNe LCs show a ``precursor bump'' prior to the main peak (e.g., \citealt{leloudas2012,nicholl2015b}; \citealt{smith2016}). This early maxima may be related to the existence of dense circumstellar media (CSM, \citealt{moriya2012}), shock breakout from an unusually extended progenitor star \citep[e.g.,][]{piro2015}, or the interaction between the SN ejecta and the progenitor's companion star \citep{moriya2015}.  

Within the magnetar scenario, \citet{kasen2016} show that precursor emission results naturally from the shock driven through the SN ejecta by the hot bubble inflated inside the expanding stellar ejecta by the magnetar wind (see also \citealt{bersten2016}).  If this ``magnetar-driven'' shock is strong enough, it becomes radiative near the stellar surface, powering an early LC bump.  This emission component is distinct from the normal shock breakout signature from the SN explosion, which occurs at earlier times and is much less luminous due to the more compact initial size of the progenitor star. 

The extremely luminous transient ASASSN-15lh \citep{dong2016} also presents a challenge to SLSN models. The peak luminosity of ASASSN-15lh exceeds that of other SLSNe by about 1 magnitude, and its total radiated energy now exceeds $3\times 10^{52}$~erg (\citealt{godoy-rivera2016}; \citealt{brown2016}). As this is near the maximum allowed rotational energy of a 1.4~\Msun\ NS, ASASSN-15lh was argued to challenge the magnetar model for SLSNe \citep{dong2016}. However, \citet{metzger2015} demonstrate that the maximum rotational energy increases with the NS mass, reaching $\approx 10^{53}$ erg for a NS close to the maximum observed mass of $\approx 2M_{\odot}$ for a range of nuclear equations of state consistent with measured NS masses and radii. Extremely luminous transients like ASASSN-15lh may indicate that some magnetars illuminating SNe can be very massive, although ASASSN-15lh itself may not only be explained by the magnetar model \citep[e.g.,][]{chatzopoulos2016} or may not even be a SLSN \citep{leloudas2016}. 

Slowly-rotating NSs can be supported against gravity only up to a maximum mass, which must exceed $\approx 2M_{\odot}$ but is otherwise poorly constrained (however, \citet{Ozel&Freire16} argue that this maximum mass is likely to be $\lesssim 2.2M_{\odot}$).  Solid body rotation can stabilize NSs with masses up to $\approx 10\%$ higher than the maximum non-rotating mass for sufficiently rapid rotation.  However, if the rotational energy of such {\it supramassive} NSs decreases below a critical minimum value (\Ecoll), then the NS will collapse to a BH on a dynamical timescale (e.g.,~\citealt{Shibata+00}).  Thus, if the magnetar produced in a core collapse SN has a mass in the supramassive range, and if it spins down to the point where its rotational energy becomes less than \Ecoll, then it will suddenly collapse to a black hole (BH) and the central energy source powering the SN will suddenly cease.  In this paper, we investigate the effect of the sudden termination of magnetar energy input due to BH transformation on the LCs of magnetar-powered SNe.

\begin{table*}
\begin{center}
\caption{Initial magnetar and SN ejecta properties of our synthetic models}
\begin{tabular}{ccccccccc}
\tableline\tableline
model & NS mass & $E_m$ & $t_m$ & \Ecoll & \tbh & $\tbh/t_m$ & \Mej &\Ni\ mass \\
     & \Msun & $10^{52}$ erg & day & $10^{52}$ erg & day & \Msun & \Msun & \Msun \\
\tableline
NS2p3m1 & 2.3 & 5.0  & 5 & 3.2 & 2.8  & 0.56 & 5 & 0.1 \\
NS2p3m2 & 2.3 & 3.5  & 5 & 3.2 & 0.47 & 0.094& 5 & 0.1 \\
NS2p4m1 & 2.4 & 12.5 & 5 & 9.3 & 1.7  & 0.34& 5 & 0.1 \\
NS2p4m2 & 2.4 & 11 & 5 & 9.3 & 0.91 & 0.18& 5 & 0.1 \\
NS2p4m3 & 2.4 & 10 & 5 & 9.3 & 0.38 & 0.075& 5 & 0.1 \\
NS2p4m4 & 2.4 & 11 & 1 & 9.3 & 0.18 & 0.18& 5 & 0.1 \\
NS2p4m5 & 2.4 & 11 & 10& 9.3 & 1.80 & 0.18& 5 & 0.1 \\
NS2p4m6 & 2.4 & 11 & 5 & 9.3 & 0.91 & 0.18& 10& 0.1 \\
NS2p5m1 & 2.5 & 17.7 & 5 & 15.4 & 0.75 & 0.15& 5 & 0.1 \\
NS2p5m2 & 2.5 & 16 & 5 & 15.4 & 0.19 & 0.040& 5 & 0.1 \\
\tableline
\end{tabular}
\label{table:magnetarproperties}
\end{center}
\end{table*}

\section{Methods}
\subsection{Energy input from magnetar spin-down}
We assume that the rotational energy of the central magnetar is emitted in a magnetized wind at the rate given by dipole vacuum or force-free spin-down (\citealt{ostriker1971}; \citealt{Contopoulos+99}). The spin-down luminosity can be expressed as
\begin{equation}
L_\mathrm{mag}(t)=\frac{E_m}{t_m}\left(1+\frac{t}{t_m}\right)^{-2},\label{eq:dipole}
\end{equation}
where $t$ is the time after the explosion, $E_m$ is the initial rotational energy of the magnetar, and $t_m$ is its spin-down timescale.

If the magnetar is supramassive and transforms to a BH after losing sufficient rotational energy, the central energy input from the magnetar suddenly ceases.  From equation (\ref{eq:dipole}), the time of BH formation (\tbh) is estimated to be
\begin{equation}
\tbh = \frac{\Delta E}{\Ecoll}t_m, \label{eq:tbhtotm}
\end{equation}
where $\Delta E\equiv E_m - \Ecoll$. Thus, the central energy input from a supramassive magnetar can be expressed as
\begin{equation}
L_\mathrm{mag}(t)=
\left\{
\begin{array}{lll}
\frac{E_m}{t_m}\left(1+\frac{t}{t_m}\right)^{-2} & & (t \leq \tbh), \\ \\
0 & & (t > \tbh).
\end{array}
\right.
\label{eq:actualinput}
\end{equation}
Although we assume that central engine activity abruptly ceases after the BH transformation, ongoing fallback accretion to the BH may in some cases provide an additional ongoing source of energy (e.g., \citealt{dexter2013,gilkis2015}; \citealt{perna2016}).

\begin{figure}
 \begin{center}
  \includegraphics[width=\columnwidth]{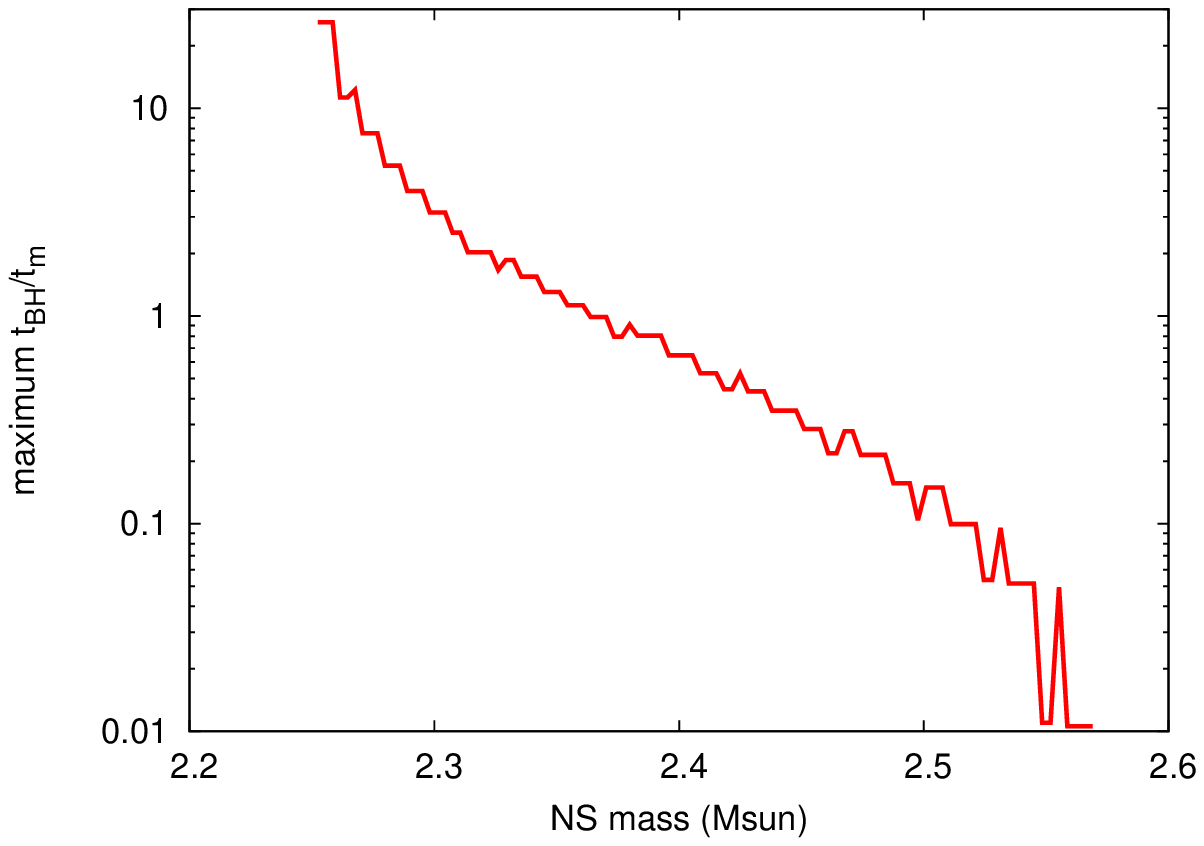}  
 \end{center}
\caption{
Maximum allowed ratio of BH formation time \tbh\ to magnetar spin-down time $t_m$, as a function of the NS gravitational mass, based on Figure~4 in \citet{metzger2015}.  The structure of the solid-body rotating NS is calculated using the {\tt rns} code (\citealt{sf95}) assuming a parametrized piecewise polytropic EOS with an adiabatic index $\Gamma = 3$ above the break density of $\rho_{1} = 10^{14.7}$ g cm$^{-3}$ at a pressure of $P_{1} = 3.2\times 10^{34}$ dyn cm$^{-2}$ (\citealt{mmb15}).  The chosen EOS results in a 1.4$~M_{\odot}$ NS radius of 10.6~km and maximum non-rotating mass of $\approx 2.24~M_{\odot}$, consistent with observational constraints.
}\label{fig:max}
\end{figure}

According to equation~(\ref{eq:tbhtotm}), there exists a maximum value of the ratio $\tbh/t_m$ due to the maximum value of $E_m$ which can be achieved for NSs of a given mass due to the mass-shedding limit \citep{metzger2015}. Figure~\ref{fig:max} shows the value of this maximum ratio as a function of the NS mass, based on Figure~4 of \citet{metzger2015}. Observe that $\tbh/t_m$ becomes $\lesssim 1$ for NSs heavier than $\sim 2.3~\Msun$.  Such massive NSs, approaching the upper allowed supramassive range, transform to BHs before losing a significant amount of rotational energy.

\subsection{Light-curve calculations}
We employ the one-dimensional radiation hydrodynamics code \texttt{STELLA} for our numerical LC calculations \citep[e.g.,][]{blinnikov1993,blinnikov1998,blinnikov2006}.
\texttt{STELLA} implicitly treats time-dependent equations of the angular moments of intensity averaged over a frequency bin using the variable Eddington method. We adopt 100 frequency bins from 1~\AA\ to $5\times 10^4$~\AA\ on a log scale. Local themodynamic equilibrium is assumed to determine the ionization levels of materials. The opacities of each frequency bin is evaluated by taking photoionization, bremsstrahlung, lines, and electron scattering into account. In particular, approximately 110 thousand lines in the list of \citet{kurucz1991} are taken into account for line opacities and they are estimated by using the approximation introduced in \citet{eastman1993}. See, e.g., \citet{blinnikov2006} for more detailed description of the code.

Starting from the initial condition described below, we deposit energy from the magnetar spin-down at the center of the exploding star. We assume that the radiation energy from the magnetar is totally thermalized (\citealt{Metzger+14}), using $L_\mathrm{mag}(t)$ (Equation~\ref{eq:actualinput}) directly as a source of thermal energy in \texttt{STELLA} \citep[cf.][]{tominaga2013}.

For comparison, we also show semi-analytic LC models from \citet{arnett1982}, which is suitable for hydrogen-free SNe \citep[cf.][]{valenti2008,chatzopoulos2012,inserra2013}. The semi-analytic LC is obtained by numerically integrating the following,
\begin{equation}
L(t)=\int^t_02\tau_m^{-2}L_\mathrm{mag}(t')t'e^{\left(\frac{t'-t}{\tau_m}\right)^2}dt'.
\label{eq:semiana}
\end{equation}
The effective diffusion time $\tau_m$ is expressed as
\begin{equation}
\tau_m = 1.05\left(\frac{\kappa_e}{\beta c}\right)^{0.5}\Mej^{0.75}\Eej^{-0.25},
\end{equation}
where $M_{\rm ej}$ and $E_{\rm ej}$ are the total mass and kinetic energy, respectively, of the initial explosion.  We assume $\kappa_e=0.1~\mathrm{cm^2~g^{-1}}$ as the electron-scattering opacity of the SN ejecta and $\beta=13.8$ \citep{arnett1982}.

\subsection{Initial SN ejecta properties}
We adopt a broken power-law structure for the initial density structure of the SN ejecta for simplicity, with a profile $\rho_\ej\propto r^{-\delta}$ at small radii which transitions to $\rho_\ej\propto r^{-n}$ outside of a break radius. Assuming homologous expansion of the SN ejecta $(r=v_\mathrm{ej}t)$, we can express the initial density structure as \citep[e.g.,][]{chevalier1989}
\begin{equation}
\rho_\ej\left(v_\ej,t\right)=\left\{ \begin{array}{ll}
\frac{1}{4\pi(n-\delta)}
\frac{[2(5-\delta)(n-5)E_\ej]^{(n-3)/2}}{
[(3-\delta)(n-3)M_\ej]^{(n-5)/2}}
t^{-3}v_\ej^{-n} & (v_\ej>v_t), \\ 
\frac{1}{4\pi(n-\delta)}
\frac{[2(5-\delta)(n-5)E_\ej]^{(\delta-3)/2}}{
[(3-\delta)(n-3)M_\ej]^{(\delta-5)/2}}
t^{-3}v_\ej^{-\delta} &
 (v_\ej<v_t), \\ 
\end{array} \right.
\label{eq:density}
\end{equation}
and
\begin{equation}
v_t=\left[
\frac{2(5-\delta)(n-5)E_\ej}{(3-\delta)(n-3)M_\ej}
\right]^{\frac{1}{2}},
\end{equation}
is the transitional velocity. We adopt $n=10$ and $\delta=1$ as typical values \citep[e.g.,][]{matzner1999}.
We adopt an initial value of $t = 10^{3}$ s in Eq.~(\ref{eq:density}).  The composition is assumed to be 50\%\ carbon and 50\%\ oxygen for simplicity.

In our fiducial model, we adopt typical properties for the SN ejecta in magnetar-powered SLSNe models of $\Mej=5~\Msun$ and $\Eej=10^{51}~\mathrm{erg}\equiv 1~\mathrm{B}$ \citep[e.g.,][]{nicholl2015}.  We fix $\Eej=10^{51}~\mathrm{erg}$ in all the models, instead varying \Mej\ to investigate the effect of the SN ejecta on the LC properties (the effects of changing \Mej\ and \Eej\ are degenerate in the LC modeling). We also place 0.1~\Msun\ of the radioactive \Ni\ at the center of the SN ejecta, although it has little effect on early LCs.

\begin{figure}
 \begin{center}
  \includegraphics[width=\columnwidth]{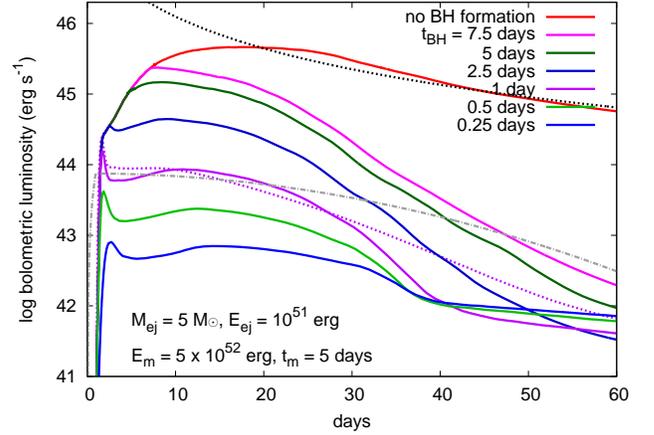}
 \end{center}
\caption{
Bolometric LCs powered by magnetar spin-down. All the models have the same input parameters except for the BH transformation time (\tbh). The solid lines are LC models obtained numerically and the dot-dashed line is the LC model obtained with the semi-analytic method assuming $\tbh = 1~\mathrm{day}$. The dotted line near the semi-analytic model is the numerical LC model with $\tbh=1~\mathrm{day}$ where the electron-scattering opacity is forced to be $0.1~\mathrm{cm^2~g^{-1}}$. The top dotted line is the input magnetar spin-down energy without BH formation.
}\label{fig:lightcurve}
\end{figure}

\section{Light curves}
\subsection{Effect of BH transformation}
Figure~\ref{fig:lightcurve} shows a series of LCs corresponding to different NS collapse times, demonstrating the effect of BH formation on the LCs of magnetar-powered SNe.  Other parameters, namely the initial rotational energy ($E_m=5\times 10^{52}~\mathrm{erg}$), spin-down timescale ($t_m=5~\mathrm{days}$), and SN ejecta properties ($\Mej =5~\Msun$, $\Eej = 10^{51}$~erg, and $\Mni=0.1~\Msun$), are held fixed in the models in Figure~\ref{fig:lightcurve}.

\begin{figure}
 \begin{center}
  \includegraphics[width=\columnwidth]{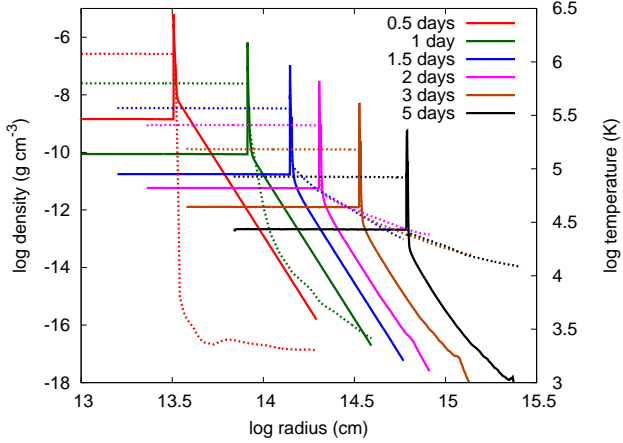}  
 \end{center}
\caption{
Hydrodynamic evolution of the numerical model with $\tbh =1~\mathrm{day}$ in Figure~\ref{fig:lightcurve}. The shock breakout occurs at around $0.5-1$~day after the explosion. The solid lines represent the density structure and the dotted lines show the temperature structure.
}\label{fig:hydro}
\end{figure}

The overall behavior of the LC evolution is reproduced reasonably well by the semi-analytic model, which in Figure~\ref{fig:lightcurve} is shown by a dot-dashed line for the same magnetar luminosity input as for the $\tbh = 1~\mathrm{day}$ model. One difference between the numerical and semi-analytic models appears in the early rising part of the LC. While the semi-analytic model shows a continuous luminosity increase from day zero, the numerical model shows an early rise and maxima starting about 1~day after the explosion. This is the effect of the magnetar-driven shock breakout described by \citet{kasen2016}.  Due to the large value of $E_m$ released by supramassive NSs, the shock is strong and radiative, and the resulting magnetar-driven shock easily reaches the stellar surface.  Figure~\ref{fig:hydro} shows the hydrodynamic evolution of the numerical model, confirming that the shock breakout indeed occurs $\approx 0.5-1$~days after the explosion. Because the shock velocity is about $10000~\kmps$ and the remaining distance from the shock to the surface is about $10^{14}~\mathrm{cm}$ at the time of the shock breakout, the photon diffusion time above the shock is about $10^{5}~\mathrm{sec}$. Therefore, the LC reaches the first maxima at about 1~day after the shock breakout.

In general, the magnetar-driven breakout bump is more prominent in the LC in cases of early BH formation. This is due to the greater peak luminosity which, for a fixed value of the spin-down time $t_{m}$, increases with the magnetar lifetime. Although the luminosity contribution due to the direct diffusion of the spin-down power eventually comes to exceed that of shock breakout, the latter persists even after this time.  

\begin{figure}
 \begin{center}
  \includegraphics[width=\columnwidth]{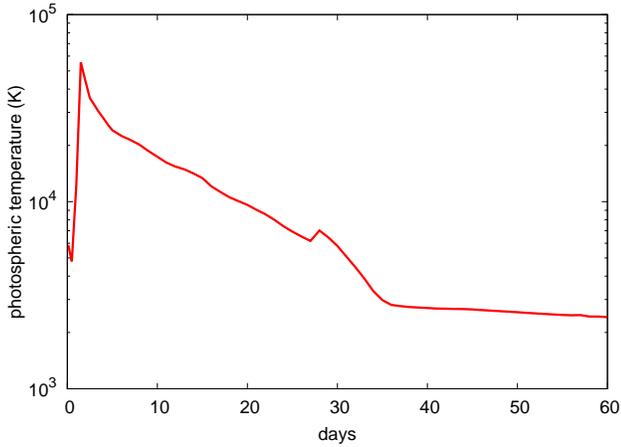}  
 \end{center}
\caption{
Photospheric temperature evolution of the numerical model with $\tbh=1~\mathrm{day}$ shown in Figure~\ref{fig:lightcurve}.
}\label{fig:photo}
\end{figure}

\begin{figure}
 \begin{center}
  \includegraphics[width=\columnwidth]{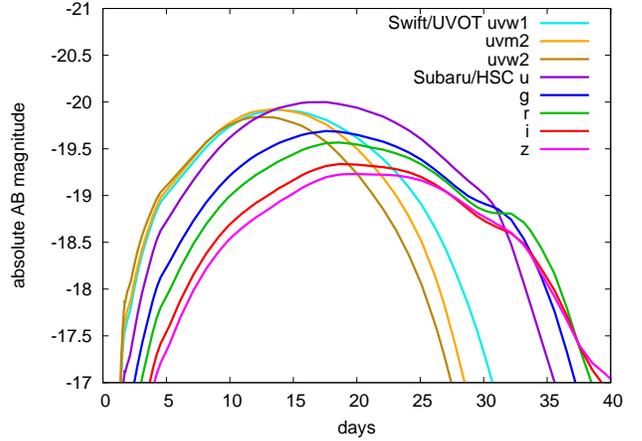}  
 \end{center}
\caption{
NUV and optical LCs of the numerical model with $\tbh=1~\mathrm{day}$ shown in Figure~\ref{fig:lightcurve}.
}\label{fig:multicolor}
\end{figure}

\citet{kasen2016} also found that the suppression of the spin-down power is required to make the LC peak due to the magnetar-driven shock breakout prominent. \citet{kasen2016} argued that the breakout peak can clearly appear if the thermalization of the spin-down energy from the magnetars is insufficient. In our model, the thermalization is kept efficient but the spin-down energy itself is shut down because of the BH transformation to make the LC peak prominent.

After the initial breakout peak, the numerical and semi-analytic LCs match reasonably well for some period of time.  However, after $t \approx$ 20~days the numerical LC begins to decline faster than the analytic expectation, presumably due to the effects of recombination and efficient adiabatic cooling in the SN ejecta. The semi-analytic model assumes a constant electron-scattering opacity of $0.1~\mathrm{cm^2~g^{-1}}$, which in reality will begin to decrease as the SN ejecta expand and cool due to recombination. For comparison, we show a numerical LC model with $\tbh=1~\mathrm{day}$ where the electron-scattering opacity is forced to be $0.1~\mathrm{cm^2~g^{-1}}$. We can see that the numerical LC with $0.1~\mathrm{cm^2~g^{-1}}$ declines slower than the actual numerical LC model, but they still decline faster than the semi-analytic model. The remaining difference likely comes from the more efficient adiabatic cooling in the numerical model with $0.1~\mathrm{cm^2~g^{-1}}$ than the semi-analytic model. A part of the magnetar energy input is used to accelerate the SN ejecta in the numerical model and the kinetic energy of the SN ejecta is increased by the magnetar. Therefore, the adiabatic cooling is more efficient in the numerical model than in the semi-analytic model where no dynamical effect of the magnetar is taken into account. In the late phases, the numerical LC tracks the decay of \Co\ decay resulting from the initial 0.1~\Msun\ of \Ni. 

Figure~\ref{fig:photo} shows the photospheric temperature evolution of the numerical model with $\tbh = 1~\mathrm{day}$, while Figure~\ref{fig:multicolor} shows the near ultra-violet (NUV) and optical LC evolution of the same model. The multicolor LCs are obtained by convolving the filter functions of Swift/UVOT ($uvw1$, $uvm2$, and $uvw2$; \citealt{poole2008}) and Subaru/HSC ($u$, $g$, $r$, $i$, and $z$; \citealt{miyazaki2012}) with the spectral energy distribution obtained by \texttt{STELLA}. Although the early shock breakout bump is clearly visible in the bolometric LC, it does not contribute appreciably to the NUV and optical bands because of the very high photospheric temperature at the time of shock breakout. The high photospheric temperature also renders the NUV and optical LCs relatively faint. Although the bolometric luminosity reaches values of $10^{44}~\mathrm{erg~s^{-1}}$ within the SLSNe range, the high photospheric temperature makes NUV and optical peak between $-19$ and $-20$~mag, below those of SLSNe.

\begin{figure}
 \begin{center}
  \includegraphics[width=\columnwidth]{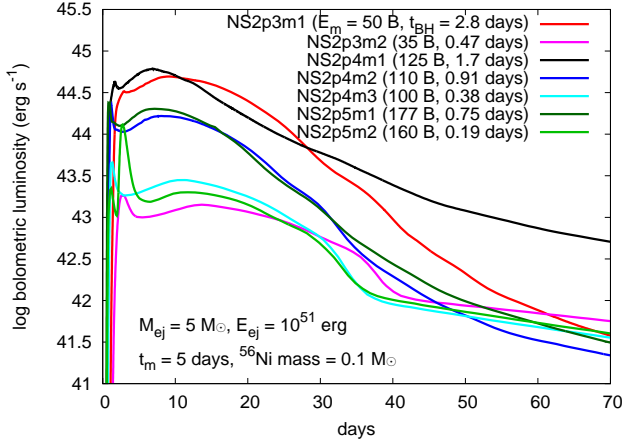}
 \end{center}
\caption{
Numerical LC models with different initial magnetar properties and BH formation times. The initial conditions are summarized in Table~\ref{table:magnetarproperties}.
}\label{fig:lightcurvediff}
\end{figure}

\begin{figure}
 \begin{center}
  \includegraphics[width=\columnwidth]{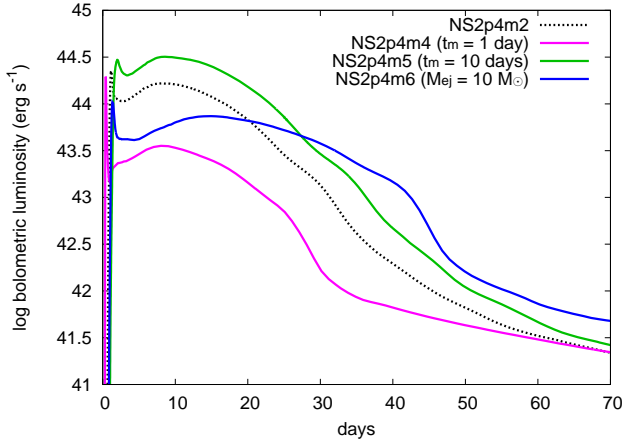}
 \end{center}
\caption{
Numerical LC models with different SN ejecta and magnetar properties (see Table~\ref{table:magnetarproperties}). The magnetar initial rotational energy ($E_m = 1.1\times 10^{53}~\mathrm{erg}$) and the NS mass (2.4~\Msun, i.e., $\Ecoll = 9.3\times 10^{52}~\mathrm{erg}$) are the same as those in NS2p4m2. The difference between NS2p4m2 ($\Mej=5~\Msun$, $\Eej = 10^{51}~\mathrm{erg}$, and $t_m=5~\mathrm{days}$) and the other models are indicated in the figure.
}\label{fig:lightcurvediffsame}
\end{figure}

\subsection{Parameter dependence}
The previous section addressed how the process of BH transformation for different formation times \tbh\ changes the LC properties of magnetar-powered SNe for fixed magnetar and SN ejecta properties. Here, we explore the effect of changing properties of the magnetar  (NS mass, $E_m$, and $t_m$) and the SN ejecta  (\Mej\ and \Ni\ mass) within their physical ranges.  For a given NS mass, there is the maximum value of $E_m$ corresponding to the mass-shedding limit \citep[e.g.,][]{metzger2015}, such that rotational energy $E_m$ must lie in the range [\Ecoll,max($E_m$)].  Thus, once $E_m$ and $t_m$ are fixed, the value of \tbh\ is no longer a free parameter (cf. Eq.~\ref{eq:tbhtotm}). We take these constraints into account in the models presented in this section, as summarized Table~\ref{table:magnetarproperties}.

Figure~\ref{fig:lightcurvediff} shows numerical LCs calculated for different $E_m$ but fixing the value of $t_m$ and the SN ejecta properties. Generally, the peak luminosity increases with higher $E_m$.  However, both the values of \Ecoll\ and max($E_m$) increase monotonically with the NS mass.  For NS with masses approaching the maximum range of supramassive NSs, the values of max($E_m$) and \Ecoll\ become sufficiently close that the NS collapses to a BH before releasing the significant amount of the rotational energy.  Extremely massive magnetars do not therefore produce bright SNe, despite the large rotational energy $E_m$ available.\footnote{The remaining rotational energy is ultimately trapped in the spin of the BH.} The maximum peak luminosity we obtain is around $10^{45}~\mathrm{erg~s^{-1}}$, which are powered by the NSs of mass $\lesssim 2.5~\Msun$ for the assumed equation of state  (Fig.~\ref{fig:lightcurvediff}). The peak luminosity ranges between $10^{43}~\mathrm{erg~s^{-1}}$ and $10^{45}~\mathrm{erg~s^{-1}}$ in our models.

Figure~\ref{fig:lightcurvediffsame} shows the numerical LCs for a fixed value of $E_m = 1.1\times 10^{53}$ erg but varying the spin-down time $t_m$ and the SN ejecta properties. Because $\tbh/t_m$ is fixed for a given value of $E_m$ and \Ecoll, then the value of $\tbh$ decreases with $t_m$.  Magnetars with shorter spin-down times $t_m$ result in smaller peak luminosities because the BH transformation occurs earlier, such that most of the magnetar energy is lost to PdV expansion instead of being released as radiation.  Models with a larger ejecta mass \Mej\ results in the longer LC duration because of the longer diffusion time, as expected.  

\begin{figure}
 \begin{center}
  \includegraphics[width=\columnwidth]{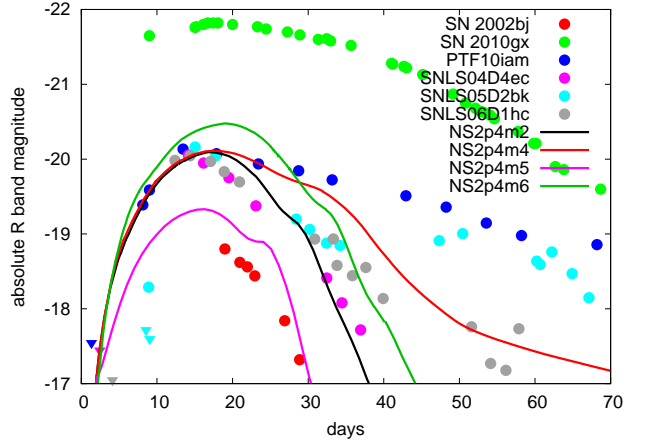}
 \end{center}
\caption{
Comparison between the synthetic $R$-band LCs and rapidly-evolving bright transients. We show the LCs of SN~2002bj \citep{poznanski2010}, a typical rapidly declining SLSN (SN~2010gx, \citealt{pastorello2010}), and the Arcavi transients (PTF10iam, SNLS04D4ec, SNLS05D2bk, and SNLS06D1hc, \citealt{arcavi2016}). The triangles in the figure indicate the upper limits of the observations.
}\label{fig:comparison}
\end{figure}

\section{Discussion}
\subsection{Comparison with observations}
Figure~\ref{fig:comparison} compares our synthetic $R$-band LCs to the measured LCs of rapidly-evolving luminous transients. As previously discussed, the optical luminosity of the synthetic LCs are relatively faint compared to the bolometric luminosity because of the high photospheric temperature. The peak luminosities of our models are typically between $-20$ and $-21$~mag. Therefore, our transients are brighter than most core-collapse SNe, yet fainter than typical SLSNe.  \citet{arcavi2016} recently reported transients precisely within this luminosity range, some of which show similar LC behavior to those predicted by our model. Although the peak luminosities of the Arcavi transients can be explained by magnetars of typical masses with the initial spin of a few ms and the magnetic field strength of $\sim 10^{15}$~G, \citet{arcavi2016} found that the overall rapid LC evolution is hard to be explained by the magnetar model. However, we accomplished the rapid evolution by shutting down the magnetar power by the BH transformation. The rapid LC evolution in our models is also consistent with that of SN~2002bj \citep{poznanski2010}.

If the \citet{arcavi2016} events and SLSNe are both powered by magnetars, then one might assume that they should occur in similar host galaxies.   However, the Arcavi transients occur in relatively higher metallicity environments than SLSNe, which instead prefer low metallicity \citep[e.g.,][]{chen2016,perley2016,leloudas2015,lunnan2015}. On the other hand, normal SLSNe are likely powered by less massive, stable magnetars, while the magnetars described in this work are necessarily very massive. Core collapse explosions giving rise to different NS masses could in principle map to different progenitor environments. Alternatively, the Arcavi transients may have several distinct origins, including those powered by magnetars transforming to BHs, and thus could originate from a diversity of environments.

We have focused on the magnetars with the spin-down timescales of the order of days which correspond to the magnetic field strengths of $\sim 10^{14}~\mathrm{G}$. Their spin-down timescales can be shorter (seconds or less) with stronger magnetic fields and such magnetars can be progenitors of, e.g., gamma-ray bursts (GRBs) \citep[e.g.,][]{metzger2015}. The BH transformation can also occur in such magnetars possibly affecting the observational properties of GRBs, but this is beyond the scope of this paper.

\begin{figure}
 \begin{center}
  \includegraphics[width=\columnwidth]{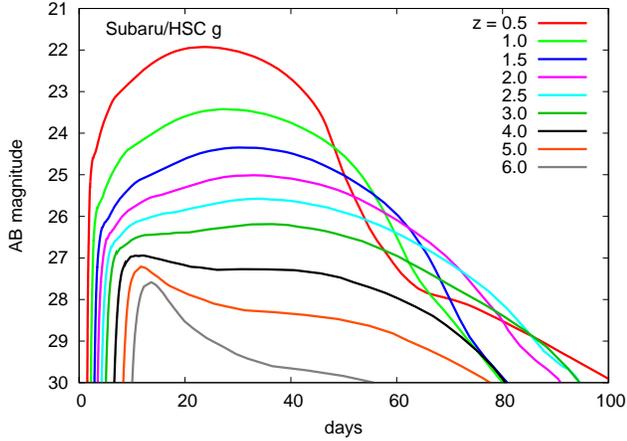}  
 \end{center}
\caption{
Redshifted LCs of the model with $t_m=1~\mathrm{day}$ in Fig.~\ref{fig:lightcurve} observed with the $g$ band of Subaru/HSC.
}\label{fig:redshifted}
\end{figure}

\subsection{Observing the magnetar-driven shock breakout bump}
Although the signature of magnetar-driven shock breakout is clearly visible in our bolometric LCs, they are difficult to observe in optical bands because of their high photospheric temperature and short duration. However, optical surveys could detect them more readily at high redshifts due to $K$-correction and time dilation effects.  Figure~\ref{fig:redshifted} shows the LCs of SN powered by magnetars transforming to BHs as observed at redshifts $z = 0.5-6$ in the $g$ band of Subaru/HSC.

The flat LC becomes visible after the initial LC rise for events at $z\gtrsim 2$, while a clear shock breakout bump appears only at $z\gtrsim 5$.  Unfortunately, a transient survey with a depth of 28~mag would be required to detect the bump in this case.  However, depending on the initial properties of the magnetar and the SN ejecta, the bump may become brighter than assumed in our fiducial models, in which case detection might still be feasible by deep transient surveys by instruments like LSST or Subaru/HSC  with a proper cadence \citep{tanaka2012}.

\section{Conclusions}
We have investigated the observational properties of SNe powered by temporarily stable supramassive magnetars which transform to BHs following a brief spin-down phase.  This sudden collapse to a BH results in an abrupt cessation of energy input from the central engine.  Our LC modeling of such transients have shown that their LCs decline much quicker than those of SN powered by the indefinitely stable, lower mass magnetars, which are usually invoked as the engines of SLSNe.  

We also find that the magnetar-driven shock breakout signal can be more significant in SNe powered by magnetars transforming to BHs, due in part to the higher rotational energy of a massive NS and the fact that prompt BH formation can allow the breakout signal to more readily shine above the normal spin-down powered LC.  Unfortunately, this breakout signal is not readily visible in NUV or optical wavebands because of the high photospheric temperature at early times.  Nevertheless, such a breakout signal could be more readily detected in optical at high redshifts, or at low redshifts by future wide-field UV transient surveys. The multi-dimensional effects like Rayleigh-Taylor instabilities in the shell causing the magnetar-driven shock breakout \citep[e.g.,][]{kchen2016} may also affect the shock breakout signatures. Our synthetic LCs of short-lived magnetars appear to be consistent with some of the rapidly-evolving bright transients recently reported by \citet{arcavi2016}.

\acknowledgments{
We thank the anonymous referee for the comments that improved this paper.
TJM is supported by the Grant-in-Aid for Research Activity Start-up of the Japan Society for the Promotion of Science (16H07413).
BDM gratefully acknowledges support from NASA grants NNX15AU77G (Fermi),
NNX15AR47G (Swift), and NNX16AB30G (ATP), and NSF grant AST-1410950, and
the Alfred P. Sloan Foundation.
The work of S.Blinnikov on development of STELLA code is supported by Russian Science Foundation grant 14-12-00203.
Numerical computations were partially carried out on the PC cluster at Center for Computational Astrophysics, National Astronomical Observatory of Japan.
}

\bibliographystyle{yahapj}

\end{document}